%% file: main.tex
\documentclass[letterpaper,twocolumn,10pt]{article}
\usepackage{usenix,epsfig,endnotes}
\input{new-commands}

\usepackage{hyperref}
\usepackage{appendix}
\newcommand{\presec}{\vspace{0in}}
\newcommand{\postsec}{\vspace{0in}}
\newcommand{\presub}{\vspace{0in}}
\newcommand{\postsub}{\vspace{0in}}

\newcommand{\Proc}{Proc. }

\newcommand{\sys}{OpenFunction}

\newcommand{\Tabs}{
  xx\= xx\= xx\= xx\= xx\= xx\= xx\= xx\=xx\= xx\= xx\= xx\=\kill}

\begin{document}

\title{\Large \bf OpenFunction: Data Plane Abstraction for Software-Defined Middleboxes}

\author{
{\rm Chen Tian$^\dagger$~~~~~~~Alex X. Liu$^\ddagger$~~~~~~~Ali Munir$^\ddagger$~~~~~~~Jie Yang$^\dagger$~~~~~~~Yangming Zhao$^\dagger$}\\
{\it $^\dagger$State Key Laboratory for Novel Software Technology, Nanjing University, China}\\
{\it $^\ddagger$Department of Computer Science and Engineering, Michigan State University, USA}\\
{Email: alexandretian@gmail.com, \{alexliu, munirali\}@cse.msu.edu}
}

\Comment{
\author{
Chen Tian$^\dagger$~~~~~~~Alex X. Liu$^\dagger$$^\ddagger$~~~~~~~Ali Munir$^\ddagger$~~~~~~~Jie Yang$^\dagger$~~~~~~~Yangming Zhao$^\dagger$\\
\affaddr{$^\dagger$National Key Laboratory for Novel Software Technology, Nanjing University, Nanjing 210023, China}\\
\affaddr{$^\ddagger$Department of Computer Science and Engineering, Michigan State University, East Lansing, MI, USA}\\
\email{alexandretian@gmail.com, \{alexliu, munirali\}@cse.msu.edu}
}
}

\maketitle
\thispagestyle{empty}

\input{alex/abstract}

\vspace{-0.015in}

{\sloppy
\input{alex/introduction}
\input{alex/related}
\input{alex/overview}
\newpage
\input{alex/actions}

\input{alex/scheduling}
\input{alex/system}

\input{alex/evaluation}
\input{alex/conclusion}
}

\newpage

\newpage
\input{alex/appendix}

\end{document}

%% file: new-commands.tex
\usepackage{cite}
\usepackage{url}
\usepackage{graphicx}
\usepackage{latexsym}
\usepackage{amsmath}
\usepackage{amssymb}
\usepackage{amsfonts}
\usepackage{comment}
\usepackage{array}
\usepackage{xspace}
\usepackage{times}
\usepackage{psfrag}
\usepackage[tight]{subfigure}
\usepackage{color}
\usepackage{algorithmic}
\usepackage[ruled,linesnumbered,vlined]{algorithm2e}
\usepackage{alltt}
\usepackage{multirow}
\usepackage{array}
\usepackage{wasysym}
\usepackage{rotating}
\usepackage{hyperref}

\newcommand{\ie}{\emph{i.e.}\xspace}
\newcommand{\eg}{\emph{e.g.}\xspace}

\newcommand{\etal}{\emph{et al.}\xspace}

\newcommand{\Comment}[1]{}

%% file: alex/abstract.tex
\begin{abstract}
%
The state-of-the-art OpenFlow technology only partially realized SDN vision of abstraction and centralization for packet forwarding in switches.
OpenFlow/P4 falls short in implementing middlebox functionalities due to the fundamental limitation in its match-action abstraction.
In this paper, we advocate the vision of Software-Defined Middleboxes (SDM) to realize abstraction and centralization for middleboxes.
We further propose OpenFunction, an SDM reference architecture and a network function abstraction layer.
%
%
%
Our SDM architecture and OpenFunction abstraction are complementary to existing SDN and Network Function Virtualization (NFV) technologies.
SDM complements SDN as SDM realizes abstraction and centralization for middleboxes, whereas SDN realizes those for switches.
%
OpenFunction complements OpenFlow as OpenFunction addresses network functions whereas OpenFlow addresses packet forwarding.
SDM also complements NFV in that SDM gives NFV the ability to use heterogenous hardware platforms with various hardware acceleration technologies.
\end{abstract}

\Comment{
The state-of-the-art OpenFlow technology only partially realized SDN vision of abstraction and centralization for packet forwarding in switches. OpenFlow/P4 falls short in implementing middlebox functionalities due to the fundamental limitation in its match-action abstraction. In this paper, we advocate the vision of Software-Defined Middleboxes (SDM) to realize abstraction and centralization for middleboxes. We further propose OpenFunction, an SDM reference architecture and a network function abstraction layer. Our SDM architecture and OpenFunction abstraction are complementary to existing SDN and Network Function Virtualization (NFV) technologies. SDM complements SDN as SDM realizes abstraction and centralization for middleboxes, whereas SDN realizes those for switches. OpenFunction complements OpenFlow as OpenFunction addresses network functions whereas OpenFlow addresses packet forwarding. SDM also complements NFV in that SDM gives NFV the ability to use heterogenous hardware platforms with various hardware acceleration technologies.
} 

%% file: alex/introduction.tex
\section{Introduction} \label{sec:introduction}
%
The vision of Software-Defined Networking (SDN) is two-fold: \emph{abstraction} and \emph{centralization}.
Abstraction means that heterogenous network devices of different vendors/architectures have common programming interfaces for implementing the functionalities needed by the control plane.
Centralization means to use a logically centralized controller in the control plane to obtain a global view of a network and to manage network resources and functionalities in an orchestrated and flexible manner.
Abstraction and centralization help SDN to achieve its two goals: \emph{breaking device vendor lock-in} and \emph{supporting network innovation}.
Abstraction helps to break device vendor lock-in by allowing third-party software to run on all SDN compliant network devices.
In contrast, traditional network devices are black boxes where both hardware and software are tightly coupled as they are typically from the same vendor, which leads to both network ossification and vendor lock-in.
Centralization helps to support network innovation by allowing network resources and functionalities to be dynamically and flexibly organized and managed.
In contrast, on traditional networks, packet forwarding and middlebox functionalities are statically deployed, which makes adopting and experimenting new network services and protocols extremely difficult.
SDN has been increasingly gaining market acceptance because of \emph{reduced cost} (by abstraction) and \emph{new functionalities} (by centralization).

Currently this SDN vision has only been realized for \emph{switches} by OpenFlow \cite{OpenFlow08Nick, OpenFlowSpecification15} and its advanced version P4 \cite{bosshart2014p4}, but not for \emph{middleboxes}.
A data communication network has two types of devices, \emph{switches} and \emph{middleboxes}.
Switches (including routers in the broader sense) provide packet forwarding.
Middleboxes provide a wide variety of networking and security functionalities such as Network Address Translation (NAT), Load Balancers (LB), firewalls (FW), WAN optimizers, proxies, IPSec gateways (VPN), and network Intrusion Detection/Prevention Systems (IDS/IPS).
For a packet, switches decide \emph{which path} the packet traverses and middleboxes decide \emph{what processing} the packet receives.

Network Function Virtualization (NFV) attempts to address the issues of tight hardware/software coupling and hardware vendor lock-in for middleboxes by implementing middlebox functions purely in software running on commodity servers \cite{etsi2012network, etsi2013network}.
Unfortunately, NFV middleboxes are software, not software-defined.
The key weakness of NFV middleboxes is low performance because commodity servers are designed for general computing purposes.
The functions and requirements of packet processing are significantly different from those of general computing.
Such differences warrant a wide variety of specialized hardware acceleration technologies such as ASIC chips and Network Processing Units (NPU), which typically have a 10-50 times performance improvement over commodity server based solutions \cite{larkins2015recent}.
This partially explains that market has been in favor of the closed-but-fast model, which uses proprietary hardware platform with fast packet processing capability, over the open-but-slow model, which uses commodity servers with slow packet processing capability.

The match-action abstractions in OpenFlow/P4 has too limited expression power to be applicable to middleboxes.
First, for the action abstraction, OpenFlow/P4 limit themselves to a fixed number of actions.
%
However, middleboxes, such as IPSec VPN, often need to perform complex payload actions such as encryption and decryption.
To address this limitation, the OpenFlow standard has kept adding new actions.
This is not a sustainable solution as packet processing actions can never be exhaustively standardized.
Furthermore, frequent standardization of new action abstraction would lead to frequent redevelopment of software/hardware, which results in vertically integrated hardware-software that must be replaced all together in network upgrading \cite{Anwer2013}.
Second, for the match abstraction, OpenFlow and P4 limit themselves to matching packet headers.
OpenFlow can only specify matching conditions over a fixed number of predefined packet header fields.
P4 improves OpenFlow by allowing users to extract customized packet header fields.
%
%
However, many middleboxes, such as application firewalls and IPSes, need to
match packet payload against signatures \cite{AlexFlowSifterJSAC14}.
Third, the table abstraction in OpenFlow/P4 is fundamentally incapable of modeling the function of scheduling.
Scheduling is essential to implement Quality-of-Service (QoS), which is further essential for many middlebox functionalities.
This inability is because the table abstraction performs splitting an aggregate flow into multiple flows whereas scheduling performs merging multiple flows into an aggregate flow.
In some sense, a table and a scheduler perform totally opposite functionalities, and one cannot model the other.

In this paper, we advocate the vision of Software-Defined Middleboxes (SDM) to realize abstraction and centralization for middleboxes.
For abstraction, an SDM architecture should have a network function abstraction layer that is \emph{platform independent} and \emph{fully programmable}.
Platform independence means to decouple the network function semantics and the underlying hardware that realizes the function.
This allows any third-party SDM program to execute at any SDM boxes (\ie, SDM compliant middleboxes) with the same semantics but different performance depending on hardware adequacy.
This abstraction layer should be hardware optimizable, which means that the performance of an SDM box can be optimized by various hardware acceleration technologies.
An SDM programmer needs not to be aware of the underlying hardware features of SDM boxes.
However, an SDM program can be translated to a platform dependent program that fully leverages the hardware features of the underlying SDM box.
Full programmability means that any middlebox functionality can be implemented by an SDM program.
For centralization, an SDM architecture should have a logically centralized controller and a suite of communication protocols that allows the controller to dynamically program the functionality of an SDM box over the network.

\begin{figure}[htbp]
\vspace{-0.12in}
\centerline{\includegraphics[width=1\linewidth]{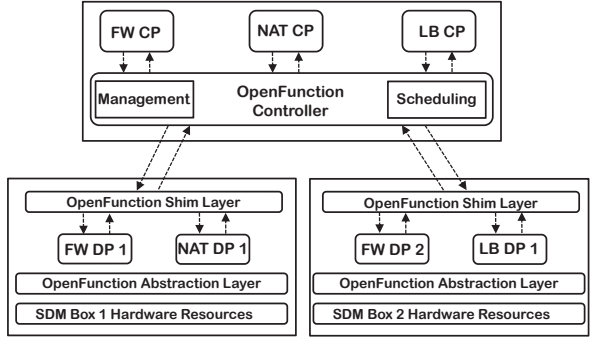}}
\vspace{-0.15in}
\caption{Software-Defined Middlebox Architecture} \label{fig:architecture}
\vspace{-0.15in}
\end{figure}

To realize the vision of SDM, we propose OpenFunction, an SDM reference architecture and a network function abstraction layer, complementing what OpenFlow achieves for switches.
Our OpenFunction architecture consists of a logically centralized \emph{OpenFunction controller} and a number of \emph{OpenFunction boxes} distributed around a network, as illustrated in Figure \ref{fig:architecture}.
Each network function, such as NAT, LB, or FW, is accomplished by a Control Plane (CP) process running on the OpenFunction controller and a group of Data Plane (DP) processes running on different OpenFunction boxes.
An DP process has a local view of the network, and its main role is to process packets in the data plane.
Based on the processed traffic, it sends predefined events to its CP process, and receives commands from its CP process.
The CP process has a global view of the network, and its main role is to make well-informed decisions for the group of DP processes that it manages.
It receives events from DP processes, performs analyses, makes decisions, and sends commands to DP processes to enforce its decisions.
The OpenFunction controller performs the management and scheduling of network resources, both CP and DP processes, and service chains.
The CP processes interact with the OpenFunction controller through its northbound API.
An OpenFunction box has a middleware layer called an \emph{OpenFunction shim layer}, which manages all DP processes running on the device, such as installs a DP, starts a DP process, and terminates a DP process.
The OpenFunction shim layer interacts with the controller through its southbound API.
A DP process interacts with the hardware resources in an OpenFunction box through an \emph{OpenFunction abstraction layer}, which provides a suite of atomic actions (\ie, APIs) for packet/flow/traffic processing.
Each DP process is implemented using these atomic operations.

OpenFunction has five categories of predefined actions:
(1) \emph{starting action}, \ie, packet sending to NICs,
(2) \emph{one-to-one actions} such as packet encapsulation or TTL decreasing,
(3) \emph{one-to-many actions} such as flow classification based on exact match or pattern-match,
(4) \emph{many-to-one actions} such as flow scheduling based on strict priority and weighted round robin,
(5) \emph{ending action}, \ie, packet receiving from hardware NICs.
For specifying customized actions beyond those predefined ones, OpenFunction has a platform-independent pseudo language.
An OpenFunction pseudo program can be compiled to platform-dependent modules for better performance.
For specifying packet processing procedures using predefined and customized actions, OpenFunction has a platform-independent scripting language.
An OpenFunction script implements a DP whereas an OpenFunction pseudo program implements an action.
Through OpenFunction abstraction, the programmer defines \emph{what to do} and the device decides \emph{how to do}.

The similarity between OpenFunction and Click modular router \cite{kohler2000click} is on the abstraction of actions.
The concept of element in Click and the concept of action in OpenFunction are similar in that they are both abstractions of packet processing operations.
The difference between OpenFunction and Click modular router \cite{kohler2000click} is two fold.
First, in Click elements, both data plane operations and control plane operations are mixed together; in contrast, OpenFunction actions contains only data plane operations.
Fundamentally, Click was not designed for SDN whereas OpenFunction is designed for SDN.
Click elements cannot be used as a data plane network function abstraction, although we do use Click elements as references in designing OpenFunction.
Second, Click is platform dependent whereas OpenFunction is platform independent.
In Click, each element is implemented by C/C++ and a program using Click library needs to be compiled before running.
In OpenFunction, pseudo programs and scripts do not need to be compiled before running; they can be distributed by the controller to OpenFunction boxes and executed at run time.

OpenFunction realizes the abstraction vision of SDN for middleboxes by the network function abstraction layer in OpenFunction boxes.
Through abstraction, for middleboxes, we decouple their hardware and software as well as their control plane and data plane.
This abstraction layer eliminates hardware vendor lock-in because the underlying hardware architecture and vendor are transparent to DP processes.
It also eliminates software vendor lock-in because any DP software can run on any OpenFunction boxes.
This abstraction layer supports hardware/software heterogeneity and promotes hardware/software innovation as the abstraction layer can be implemented in a variety of hardware/software technologies such as ASIC, multi-core, FPGA, and GPU.
This lowers the middlebox market barrier because as long as a hardware device supports the abstraction layer, it can immediately support DP processes and thus can be immediately deployed on existing networks to work with existing devices.
This abstraction layer also significantly improves the programmability of middlebox functionalities because a DP developer does not need to implement those common operations in packet processing.
From the software engineering perspective, this abstraction layer promotes software reuse.
This will significantly shorten the time-to-market cycle.

OpenFunction realizes the centralization vision of SDN for middleboxes by the logically centralized OpenFunction controller.
As it is well understood that centralization in OpenFlow brings a long list of benefits to packet forwarding such as flexible path selection, better load balancing, finer-grained network control, less configuration errors, higher manageability, and increased reliability and security, centralization in OpenFunction brings many benefits to network functions.
For example, the global view that a CP process obtains from the events generated by the group of DP processes that it manages helps the CP to make better informed decisions, \eg, an IDS CP process can perform correlation analysis on the events from its DP processes to better identify attack activities.
For another example, the ability that a CP process can issue commands to its DP processes allows the CP to orchestrate the DP processes to collaboratively accomplish some tasks.

OpenFunction, together with OpenFlow, makes service chain construction and scheduling much easier than the current NFV architecture.
A service chain is a sequence of middlebox functions that the network operator wants certain traffic to traverse in order.
To construct and schedule service chains, the OpenFunction controller performs resource allocation according to a given optimization policy (\eg, load balancing) in collaboration with the forwarding plane authority (\eg, an OpenFlow  controller).
For example, suppose the network operator wants to allocate 10G bandwidth between two networks and also all traffic between the two networks to pass through an IDS'', first, the OpenFlow controller finds an appropriate path between the two networks where each switch on the path has at least 10G bandwidth and the path contains at least one OpenFunction box, second, the OpenFunction controller starts a new IDS DP process on that OpenFunction box.
In this SDM+SDN and OpenFunction+OpenFlow paradigm, one technical challenge is service chain scheduling: given a service chain that a flow needs to traverse, the controller chooses a sequence of middlebox data plane processes (running on one or more OpenFunction boxes) to form the service chain.
The scheduling needs to satisfy the three requirements of service chain constraints, device resource constraints, and flow conservation.
Furthermore, it needs to be done in a load balancing way to minimize the maximum utilization of the resources in candidate devices.
To address this challenge, in this paper, we formulate service chain scheduling as a NP-Hard problem, and proposed both offline and online scheduling algorithms.

OpenFunction, together with OpenFlow, helps to support the vision of service oriented network management.
While OpenFlow makes switches easier to manage, currently middleboxes are still notoriously difficult to manage because the hardware and software of those middleboxes are often tightly coupled and different middleboxes of different vendors often have different management interfaces.
Service oriented network management means to automate the translation from high-level service oriented network management requirement to low-level network device configurations.
For example, a high-level service oriented network management requirement can be to ``allocate 10G bandwidth between two networks and to ensure that all traffic between the two networks to pass through an IDS''.
While translating this high-level requirement to network device configurations in today networks is a complex, laborious, and error prone process, it can be easily done in our proposed OpenFunction+OpenFlow architecture as a service chain.

Our SDM architecture and OpenFunction abstraction are complementary to existing SDN and NFV technologies.
SDM realizes abstraction and centralization for middleboxes, whereas SDN realizes those for switches.
SDM breaks device vendor lock-in and supports network innovation for middleboxes, whereas SDN does those for switches.
OpenFunction addresses network functions whereas OpenFlow addresses packet forwarding.
SDM complements NFV in that SDM gives NFV the ability to use heterogenous hardware platforms with various hardware acceleration technologies.

We implemented a working SDM system including one OpenFunction controller and three OpenFunction boxes based on Netmap, DPDK and FPGA respectively.
We also develop two stateless network functions (\ie, NAT and IPsec), two stateful network functions (stateful firewall and IPS).
We draw the following conclusions from our experimental results.
First, middlebox functions implemented using OpenFunction abstraction can achieve high performance.
For example, our FPGA platform can achieve near 10G bps throughput for NAT and IPSEC middleboxes.
%
%
%
Second, the performance of OpenFunction scripts is platform dependent.
Third, for service chain scheduling, our offline algorithm can achieve near to optimal load balancing whereas our online algorithm is fast but the performance is suboptimal.
Fourth, the overhead of enforcing service chain both across and inside OpenFunction boxes is negligible.

\Comment{ 
The rest of paper proceeds as follows.
We first review background and motivation in Section~\ref{sec:related}.
The overview of our design is presented in Section~\ref{sec:overview}.
Then, we introduce our middlebox data plane abstraction in Section~\ref{sec:actions}, and control plane service chain scheduling
algorithms in Section~\ref{sec:scheduling}.
The implementation of the prototype system is described in Section~\ref{sec:system}.
Extensive evaluation results are demonstrated in Section~\ref{sec:evaluation}.
We conclude our paper in Section \ref{sec:conclusions}.
}

%% file: alex/related.tex
\section{Background and Related Work} \label{sec:related}

\noindent
\textbf{Router Abstraction:}
Some prior work is on designing an abstraction layer to support portability across heterogeneous platforms.
Handley \etal built extensible routers on top of commodity platforms in XORP, focusing on breaking the complex control plane of routing protocols \cite{handley2003xorp,handley2005designing}.
Mogul \etal advocated an abstraction layer called Orphal to support portability of third-party software in a position paper \cite{mogul2008api}.
XORP and Orphal significantly differ from OpenFunction.
First, XORP and Orphal are not designed for SDN whereas OpenFunction is.
Second, XORP and Orphal are not platform independent whereas OpenFunction is.
Furthermore, XORP focuses on routing whereas OpenFunction focuses middlebox functionalities, and Orphal is based on router specific hardware resources (such as TCAM and DPI engines) whereas OpenFunction is based on middlebox functionalities and is hardware agnostic.
Song proposed a hardware abstraction called POF, which focuses on protocol-oblivious forwarding \cite{song2013protocol}.
It extends the OpenFlow instructions with protocol-oblivious operations such as \emph{AddField} and \emph{SetFieldFromValue}.
%
%
Answer \etal propose to decouple the processing at the controller and at middleboxes while providing an interface by which they can communicate \cite{anwer2013slick}.
This vision is similar to us; however, neither design nor implementation details were given in the two-page position paper \cite{anwer2013slick}.

\noindent
\textbf{Data Plane Programmability:}
Several systems focus on the data plane throughput of extensible software-based routers.
These systems are orthogonal to OpenFunction and they are useful to build OpenFunction boxes.
RouteBricks exploits parallelism both across multiple servers and across multiple cores inside each single server ~\cite{dobrescu2009routebricks}.
PacketShader exploits the massive parallelism of GPU for batch processing \cite{han2011packetshader}.
GASPP focuses on stateful packet processing, and its GPU-accelerated framework can achieve multi-gigabit processing \cite{vasiliadis2014gaspp}.
Some systems focus on data plane extensibility of new network forwarding protocols.
The main direction is to pair the high-throughput ASIC processing path with a fully programmable co-processer to enhance the programmability.
ServerSwitch uses x86 CPU as the co-processor \cite{lu2011serverswitch, lu2012using}.
SSDP instead uses an NPU-driven subsystem \cite{narayanan2012macroflows}.
SwitchBlade allows rapid prototyping in FPGA \cite{anwer2011switchblade}.
OpenDataPlane is an open-source cross-platform set of APIs for programming network processors \cite{OpenDataPlane}.
OpenDataPlane abstraction is at a lower level than OpenFunction as OpenDataPlane specifies \emph{how to do} by directly manipulating device resources whereas OpenFunction only specifies \emph{what to do} and let each device decide how to implement each action.

%% file: alex/overview.tex
\begin{figure*}[htbp]
\centerline{\includegraphics[width=1\linewidth]{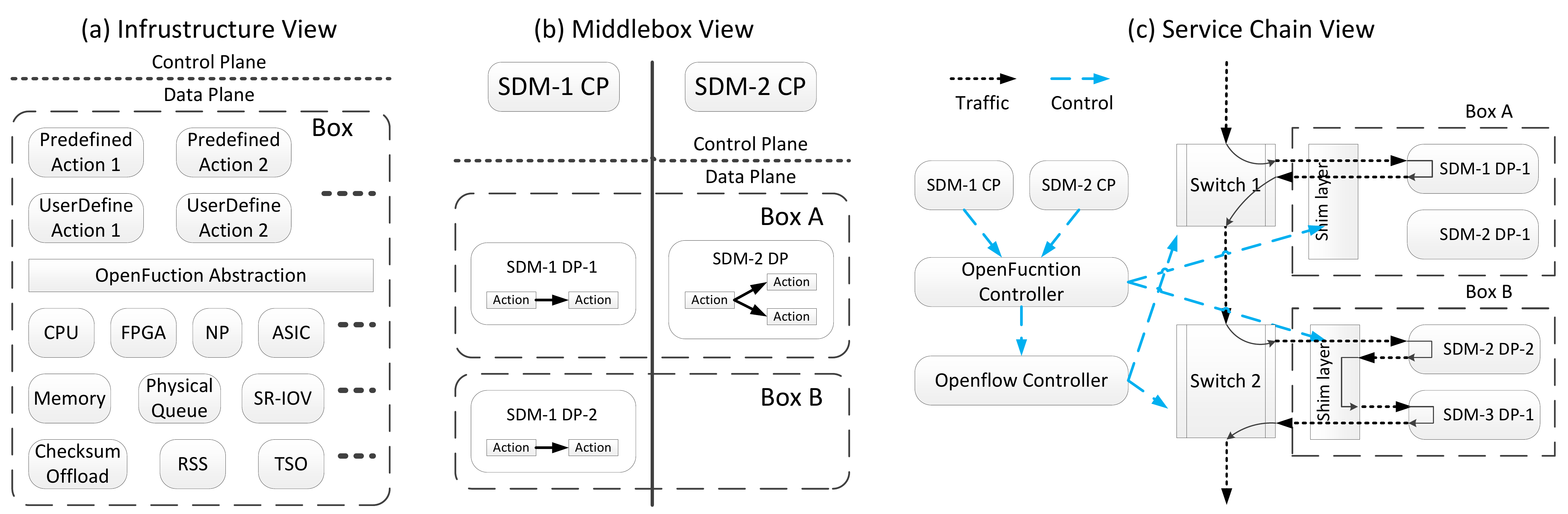}}
\vspace{-0.18in}
\caption{OpenFunction Architecture Overview} \label{fig:architecture}
\vspace{-0.18in}
\end{figure*}

\section{OpenFunction Architecture Overview} \label{sec:overview} \postsec
OpenFunction architecture consists of a logically centralized OpenFunction controller and a number of OpenFunction boxes distributed around a network where each box has a network function abstraction layer.
Here we give an overview of OpenFunction from three perspectives: the data plane, middleboxes, and service chains.

\noindent\textbf{Data Plane View:}
OpenFunction exposes an extensible set of actions to the control plane as shown in Figure \ref{fig:architecture} (a).
The semantics of an action is to take a packet as its input, perform some operations, and either push the packet to the next action or
wait for the next action to pull the packet.
OpenFunction is action oriented, similar to the object oriented programming paradigm.
Each action object is an instance of an action class.
An action class is a self-contained and functionally independent packet processing unit, such as decreasing the TTL field and calculating the TCP checksum.
Each action class consists of some data members, which we call attributes.
For example, a \texttt{PacketCounter} action may need to update its counter attribute whenever a packet passes through this action.
An action may also send an event to the control plane.
For example, a \texttt{TCPSyncNotifier} action may send a TCP sync event whenever a TCP sync packet passes through it.
There are two kinds of actions: pre-defined actions and user-defined actions.
Pre-defined actions are those supported by OpenFunction compliant boxes and user-defined actions are those written by users using a platform-independent pseudo language.
We allow multiple DP processes running on the same OpenFunction box, similar to router visualization, where in the same box, different DP processes belongs to different CP processes for different middlebox functions.

\noindent\textbf{Middlebox View:}
In OpenFunction, a network function (such as NAT, LB, or FW) is implemented by a Control Plane (CP) process and a set of Data Plane (DP) processes running on OpenFunction boxes, as shown in Figure \ref{fig:architecture} (b).
A DP is modeled as a directed acyclic graph where each node is an action.
A DP process has a local view of the network and its main role is to process data plane packets.
Based on the processed traffic, it may sends events to its CP process, and receives commands from its CP process.
The CP process has a global view (via the controller) and its main role is to make well-informed decisions for the group of DP processes that it manages, according to its middlebox functionality.
It receives events from DP processes, performs analyses, makes decisions, and sends commands to DP processes to enforce its decisions.
We call a CP process together with its group of DP processes a \emph{Software-Defined Middlebox} (SDM).
With OpenFunction, implementing a middlebox functionality on the data plane becomes much simpler as it mostly involves composing a graph of pre-defined actions, possibly with a few user-defined actions.
Thus, SDM developers mostly focus on designing CP programs, which are often the most innovative part of their implementation.

\noindent\textbf{Service Chain View:}
A service chain is a series of middlebox functions that a packet needs to go through, as shown in Figure \ref{fig:architecture} (c).
The OpenFunction schedules and enforces service chains.
The controller monitors resource (such as CPU, memory, or a specific acceleration card) utilization and executes an service chain scheduling algorithm; it then coordinates with the forwarding plane authority (\ie, an OpenFlow controller) to enforce the chain.
Each OpenFunction box has a middleware layer called an OpenFunction shim layer, which manages all DP processes running on the device, such as installs a DP, starts a DP process, and terminates a DP process.
The shim layer is fully controlled by the controller and the controller monitors the resource unitization through it.
Note that service chains needs to be enforced both across and inside boxes.
An Openflow controller still controls the forwarding between boxes.
The OpenFunctoon controller sends flow steering instructions to the Openflow controller.
A dedicated IO process of the shim layer steers flows among functions according to their service chains.
Figure \ref{fig:architecture} (c) shows a service chain $SDM~1\rightarrow SDM~2 \rightarrow SDM~3$.

%% file: alex/actions.tex
\presec \section{Abstraction Design} \label{sec:actions} \postsec

\presub \subsection{Abstraction Specification} \postsub
\label{sec:actions:specification}
\sys~uses a script for dataplane specification: an example of IPSEC middlebox data plane is shown in Algorithm~\ref{code:ipsec}.
%
%
Hereby is the brief description of \sys~script: 
each action instance definition line starts with a ``@'';
each line defines a connection between an egress port of one action and an ingress port of another action;
action definitions and connection definitions together construct the action graph of a DP process.

%
%
%


%
To carefully tradeoff composability and optimizability, we analyze a number of middleboxes such as NAT, IPSec, and IDS, and design a suite of atomic actions that we classify into the following five categories:
\begin{itemize}
\item \emph{starting action:} These actions send packets to either hardware/virtualized NICs, or specific memory locations (\eg, FromDevice in the example ).

\item \emph{one-to-many action:} This category consists of actions based on the following four matches: exact-match (\eg, used in the example), longest-prefix-match (\eg, used in FIB), first-match (\eg, used in ACL), and pattern-match (\eg, used in DPI).

\item \emph{many-to-one action:} A flow scheduling action takes multiple flows as input, and outputs one combined stream of traffic. Example flow scheduling operations include strict priority (SP) and weighted round robin (WRR).

\item \emph{one-to-ones action:} These actions perform a single packet processing such as changing the length of a packet, either at the head (\ie, Encap/Decap) or the tail (\ie, Pad/Unpad), and modifying a field or the metadata of a packet (\ie, decreasing the IP TTL value and setting the ECN bit are two actions).
    
   \item \emph{ending action:} These actions receive packets from either hardware/virtualized NICs, or specific memory locations (\eg, ToDevice in the example).
\end{itemize}
The attributes of an action can be set either at the configuration phase, or at run-time.
Stick to the IPSEC example, for a given IPSEC tunnel, the IP\_SRC/IP\_DST attributes of ChangeSrcIP/ChangeDstIP actions are set at the configuration phase as parameters.
While for the ESPEncap action, its security index SPI and replay parameter RPL should be negotiated by the control plain at run-time;
the IPSEC DP process leaves the two attributes unset at first, and lets the CP process performs negotiation before setting the attribute value.

As the target of action operations, a packet class consists of three types of data: fields, metadata, and properties.
The metadata is a piece of memory associated with a packet and is used to store modifiable information about the packet, such as the ID of the physical port from which the packet is received.
The properties means some non-modifiable information about the packet, such as the packet size.
Although some properties of a packet may change after certain actions, \eg, the length of a packet changes after Encap/Decap, the properties of a packet cannot be written or modified by action functions directly.
Table \ref{table:middleboxes} summaries the accessibility of packet data.

\begin{table}[htb]
\begin{center}
\scalebox{0.85}{
\begin{tabular}{|c|c|c|c|c|c|}
\hline Targets & \multicolumn{3}{|c|}{Packet} & \multicolumn{2}{|c|}{Action} \\
\hline & Field & Metadata &  Property &  Attributes & Event \\
\hline
Read  & Yes &  Yes  &  Yes &  Yes  &  No \\
\hline
Write & Yes &  Yes  &  No  &  Yes  &  Yes \\
\hline
\end{tabular}}
\vspace{-0.1in}
\caption{Summary of data access}\label{table:middleboxes}
\vspace{-0.15in}
\end{center}
\end{table}

To prevent the exact fate of OpenFlow's vertical integration, we do not enforce any standard action.
Instead, a set of actions are \emph{recommended}; and for each such action, a default implementation written in \sys~pseudo language (next part) is provided together,
An \sys~box can flexibly choose the action to optimize by exploiting their underlying hardware acceleration capabilities.
For the actions whose optimized implementation is not available at the box, the default pseudo code can be used, like user-defined actionS.

Furthermore, we allow \sys~boxes to provide multiple implementations of the same action: for example, one version is CPU based and another version is GPU based.
The controller can dynamically determine which version to load/unload based on the availability of network resources and the processing requirements of
flows at run time.
Later in Section~\ref{sec:scheduling}, we consider this flexibility in formulating and solving the service chain scheduling problem.

\presub \subsection{Pseudo Language} \postsub
\label{sec:actions:language}
We propose a platform-independent pseudo language that allows SDM developers to design arbitrary new middlebox data plane action, and supported by any \sys~enabled device.
The vendor of an \sys~box needs to provide a source-to-source translator, which takes a pseudo C program as its input and outputs a native program that exploits the hardware acceleration strength of the box.
The output platform-dependent program is further compiled into executable components using the box's native compiler.
As a proof-of-concept, shown in Section~\ref{sec:system}, in our prototype system an action class can be translated either to a platform-dependent C/C++ program such as a Click element, or a FPGA block (in Verilog).

\begin{figure}[htbp]
\centerline{\includegraphics[width=1\linewidth]{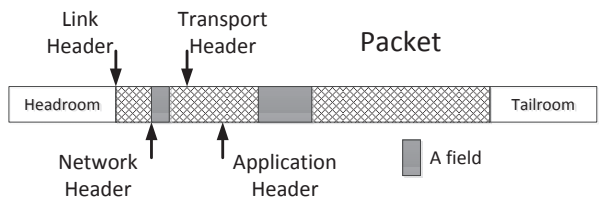}}
\caption{Abstraction of packet field} \label{fig:data}
\end{figure}

We now introduce the packet field abstraction that the pseudo programs operate on.
The whole packet is assumed to be stored in a continuous memory location as shown in Figure \ref{fig:data}.
There is large headroom/tailroom before/after the packet data.
Such spaces are reserved for actions that change packet length (\ie, Encap/Decap and Pad/Unpad).
We also assume the existence of a mandatory preprocessing step, which marks the start offset of link/network/tranport/app layers (if they exist).
Each field is a continuous range of bytes in a packet memory range, defined by a tuple \texttt{(header, offset, type)} and can be attached with a name.
For example, IP checksum can be defined as a field starts at the 10th bytes offset of the network header by the \texttt{FIELD} keyword.
The length is automatically determined by its data type \texttt{UNIT16}.
Note that a special type is \texttt{DATA}, which represents a raw array of bytes.

The metadata of a packet is accessed as a named object (with keyword \texttt{META}).
In OpenFunction, a \texttt{(type, length)} tuple defines a specific metadata object.
The \texttt{length} configuration is required only when the \texttt{type} is \texttt{DATA}.
This design provides the system with the freedom to arrange a meta object in its metadata memory.
Packet properties (with keyword \texttt{PROP}) and action attributes (with keyword \texttt{ATTR}) are also accessed as named objects.
Note that packet properties are limited to a small set of intrinsic packet characteristics and read-only to program.
The additional feature of an attribute is that it can be read/written by another action or the control plane.
Each event object is a continuous raw array of bytes.
It is up to the action developer to decide the content of the event.

A key design decision is to separate packet access (Table \ref{table:middleboxes}) codes and normal codes.
More specifically, we use character "@" at the start of a line to denote a data access line.
Following the "@" letter, there should either be a packet variable definition of \texttt{FIELD}, \texttt{META},  \texttt{PROP} or \texttt{ATTR}, or \texttt{LOAD}/\texttt{STORE} to access the data.
Currently, we do not support direct algorithmic operations over packet data variables; they are used to denote the physical locations in the packet.
Instead, we first load them into local variables, perform calculation, and then store them back.
The benefit is that all non-data-access codes can be written as normal C codes.

Algorithm \ref{alg:SetIPChecksum} shows an example pseudo program.
This user-defined \emph{SetIPChecksum} action accepts an incoming IP packet and calculates and sets its IP checksum field; here the action assumes that the incoming packets are with MAC headers.
As an example, the \emph{iphlen} is defined first (Line 1), and loaded into a local variable \emph{hlen} before used (Line 5).
%


%% file: alex/scheduling.tex
\presec \section{Service Chain Scheduling} \label{sec:scheduling} \postsec
One challenging task for \sys~controllers is to perform service chain scheduling: given a flow and a sequence of SDM functions that the flow needs to pass through, the controller needs to choose a sequence of DP processes of that SDM so that the flow can go through this sequence of DP processes.
Note that if a specific action in the program has multiple implementations (\eg, both CPU-based and GPU-based), then the compiled SDM data plane also has multiple implementations.
The controller can load any desired implementation according to the resource scheduling and flow requirements.
For example, if a middlebox's data plane graph has 3 actions and each action has 2 implementations; in total there could be $2^3=8$ implementations stored for this middlebox at the maximum.
Due to the cheap price of storage, we consider this exponential increase in the number of implementations a trivial concern.

There are some related work in service chain scheduling and enforcement.
CoMb exploits consolidation opportunities in NFV deployment to save cost \cite{sekar2012design}.
It also provides solutions to resource management, traffic redirection, and coordination among different data plane processing elements.
However, it requires that the middlebox processes of the whole service chain pertaining to a given session flow run on the same device.
This is not optimal in many \sys~scenarios: image that one hardware platform has special optimization for IPSEC, and another platform has special optimization for cache; force the whole service chain in a single node compromises the benefits of global resource optimization.
Some schemes have been proposed to solve the problem of steering traffic among physical middleboxes to follow the network policy (\ie, service chain).
As a pioneer work, the pswitches (\ie, policy-aware switches) scheme adds a new layer-2 for data centers, which can steer the network traffic through unmodified middleboxes \cite{joseph2008policy}.
Using SDN, SIMPLE can enforce flow-level policy even if middleboxes modify the packets or change the session level semantics \cite{qazi2013simple}.
FlowTag makes a further step by attaching tags to flow packets to enable flow tracking.
These existing schemes can be used in \sys~to control the service chain among middlebox platforms \cite{fayazbakhsh2014enforcing}.

%
Table \ref{table:symbols} summarizes the symbols used in this section.
For a given flow, the OpenFunction controller needs the following three types of inputs to perform service chain scheduling.
\begin{itemize}
\item \emph{Chaining Requirement:} Each flow has a classification rule that uniquely identifies the flow and estimated bandwidth. For the given flow, we use $SDM_1, SDM_2, \cdots, SDM_K$ to denote the sequence of $K$ SDMs that it needs to traverse. This flow specification is typically specified by network operators.
\item \emph{Implementation Availability:} For each $SDM_k$, let $V_k$ denote the total number of different implementations of $SDM_k$. Each implementation has its specific resource requirement, which is usually multi-dimensional as different actions in an implementation require different resources.
\item \emph{Box Capability:} Let $N$ denote the set of \sys~boxes. Let $\Omega_{n,r}$ denote the capabilities of resource $r$ in each box $n$. Also, the existing utilization of resource $r$ in OpenFunction box $n$ is denoted as $\Gamma_{n,r}$ (in percentage). The OpenFunction controller obtains box capabilities by performing resource monitoring of each OpenFunction box in real time.
\end{itemize}
As a practical constraints, all packets in a given flow should traverse the same DP process for a given SDM in the chain.

\begin{table}[htbp]
\caption{Symbols used in this paper} \label{table:symbols}
\centering
\begin{tabular}{|p{0.8cm}|p{6.3cm}|}
\hline $k, K$ & SDM k, $1\leq~k\leq~K$. \\
\hline $n, N$ & OpenFunction box n, $1\leq~n\leq~N$. \\
\hline $r, R$ & resource r, $1\leq~r\leq~R$. \\
\hline $N_{kl}^f$ & Input binary constant of flow $f$'s service chain to indicate if SDM $l$ is the next SDM of $k$\\
\hline $D_{k,i,r}$ & The amount of resource $r$ is required to support a unit of flow for SDM $k$ with implementation $i$. \\
\hline $\Gamma_{n,r}$ & the existing utilization of resource $r$ in box $n$. \\
\hline $\Omega_{n,r}$ & the capabilities of resource $r$ in box $n$. \\
\hline $A_f$ & The amount of flow $f$\\
\hline $Z_{k,i,n}^{l,j,m,f}$ & Binary variable to indicate if flow $f$ is traversed from the SDM $k$ with implementation $i$ on box $n$ to SDM $l$ with implementation $j$ on box $m$.\\
\hline $X_{k,i,n}$ & The amount of flows traversing node $n$ for SDM $k$ with implementation $i$. \\
\hline
\end{tabular}
\end{table}

\noindent\textbf{ILP Formulation}
The optimization objective in service chain scheduling is load balancing across all resources of all nodes.
\begin{equation} \label{equ:obj}
\text{minimize}\ \ \ \ \ \ \ \ \ \ \max \frac{\sum_k \sum_i X_{k,i,n}D_{k,i,r}}{\Omega_{n,r}}+\Gamma_{n,r}
\end{equation}

The main constraint is the flow conservation for each flow crossing the SDMs
\begin{equation} \label{con:flowconservation}
\sum_{l,j,m\geq n}Z_{k,i,n}^{l,j,m,f} - \sum_{l,j,m\leq n}Z_{l,j,m}^{k,i,n,f} = \left\{ \begin{array}{ll}
1 & \textrm{if $k=0,n=0$}\\
-1 & \textrm{if $k=K,n=N$}\\
0 & \textrm{otherwise}
\end{array} \right.
\end{equation}
In addition, a flow may traverse from SDM $k$ to SDM $l$ only if SDM $k$ is followed by SDM $l$ on flow $f$'s service chain,
\begin{equation} \label{con:nextHop}
N_{kl}^f \geq Z_{k,i,n}^{l,j,m,f}
\end{equation}
Based on this flow conservation constraint, the traffic can be calculated by
\begin{equation} \label{con:demandCalcu}
X_{k,i,n}=\sum_{f}\sum_{l,j,m}A_f Z_{l,j,m}^{k,i,n,f}
\end{equation}
To derive an ILP formualation, we can introduce an auxiliary variable $z$ to present the objective function and $z$ should satisfy
\begin{equation} \label{con:auxVariable}
z \geq \frac{\sum_k \sum_i X_{k,i,n}D_{k,i,r}}{\Omega_{n,r}}+\Gamma_{n,r}
\end{equation} \label{equ:finalObj}
and the objective becomes
\begin{equation}
\text{minimize}\ \ \ \ \ \ \ \ \ \  z
\end{equation}
Therefore, the Integer Linear Programming (ILP) formulation of this problem can be summarized as the resource utilization optimization problem (RUOP).
Obviously, the ILP model formulated above is intractable in large size network.
Specifically, due to the large dimension of binary variable $Z_{k,i,n}^{l,j,m,f}$, there will be more than $10^7$ binary variables in the model even if each index has only 10 possible values.
Accordingly, we need to design an efficient heuristic to solve this problem.

\noindent\textbf{Offline Algorithm}
We first consider the offline scheduling, where a number of flows are given and can be optimized at one time.
Since the most critical element which makes our problem intractable is the large number of binary variables, we consider relaxation and rounding method to solve this problem.
For clear presentation, we call the enforced service chain (\ie, the ordered SDM realized by which box and with which implementation, as the path of this flow.
Following the relaxation and rounding idea, we first relax the binary constraint of the variable $Z_{k,i,n}^{l,j,m,f}$, \ie, treat it as a real variable in the range $[0,1]$, and solve the derived Linear Programming model.
With this solution, each flow may be split among multiple paths which are determined by the value of $Z_{k,i,n}^{l,j,m,f}$.
To get unsplitable path for each flow, a simple method is to directly round each flow to the path which carries the largest fraction of this flow.
However, this method cannot obtain a good result especially when there are a lot of boxes and many implementation versions for each SDM.
In this case, each path may carry only a little fraction of the flow; directly rounding each flow to the path carrying maximum fraction of flow may deviate too far away from the ILP optimum.
To solve this problem, we propose a progressive method for the rounding phase to get unsplitable paths.

Respect to progressive rounding, the main idea is to iteratively forbid some links that carrying very little fraction for each flow, till there is only one path for each flow or there is a path carrying most of the fraction of flow.
Based on above discussions, we design Algorithm \ref{alg:offline} to solve the RUOP formulation.
In this algorithm, there is one step that has not been discussed.
In Line \ref{lin:threshold}, we should find a threshold $\epsilon$ to determine if a $Z_{k,i,n}^{l,j,m,f}$ should be set to 0 for the rounding purpose.
This $epsilon$ can be set based on the tradeoff between the algorithm performance and the algorithm running time. With large $epsilon$, more $Z_{k,i,n}^{l,j,m,f}$ will be set to 0 in each loop and hence there are fewer loops executed in the algorithm.
With smaller $epsilone$, we can do derive a better solution.
On the other hand, we should update the $epsilon$ in each loop since the $Z_{k,i,n}^{l,j,m,f}$ should become larger in each loop as there are fewer paths can be used by the flows with the algorithm execution.
Accordingly, without updating the rounding threshold, there will be endless loop in the algorithm.

\noindent\textbf{From Offline to Online}
We can also extend the algorithm to online version of this problem, because in reality
service chain requests may arrive randomly, not all at once.
The first scheme is to hold the flows for a while to get a bulk of flows and then trigger the offline algorithm to reserve SDM resources for these flows.
If the flow arriving rate is very small, hold flows to wait for the later flows may be a large overhead for the flow arriving earlier.
In this case, we can treat each flow as a bulk to trigger Algorithm \ref{alg:offline} to set up SDMs for each single flow.

%% file: alex/system.tex
\vspace{-0.3in}
\presec \section{Prototype Implementation} \label{sec:system} \postsec
\presub \subsection{Platform} \postsub
We build a proof-of-concept system to verify the \sys~abstraction and get an estimation of achievable performance.
%

\noindent\textbf{Data Plane~}
We develop two x64 platform \sys~boxes based on DPDK and Netmap, and one hardware box based on FPGA.
Netmap is a tool designed for high speed packet I/O in commodity servers.
Implemented as either a modified NIC driver or a kernel module, Netmap also enables high-speed processing in user space.
Supported by Intel, Data Plane Development Kit (DPDK) is a combination of data plane libraries and NIC drivers for fast packet processing in user space.
Poll mode drivers (PMD) are designed to enable direct packets exchange between a user space process and a NIC.
A low overhead \emph{run-to-completion} model is used to achieve fast data plane performance.
In our testbed, there are 16 Dell R320 machines, each has a 10G Intel X520-SR1 ethernet NIC and is both DPDK and Netmap enabled.
We also develop a prototype box in a FPGA card via Xilinx Vivado High-Level Synthesis tool, which supports C/C++ to describe designs and translates them into hardware description languages (HDL) such as VHDL or Verilog HDL~\cite{meeus2012overview}.
The platform is a programmable FPGA card with a Xilinx Kintex 7 chip, two 10 Gbps NIC and four 1 Gbps NIC.

For each prototype box, we need to provide three support: \sys~script, \sys~language and \emph{in-box chaining}.
Regards the third requirement, the device needs to support dynamic DP function loading/unloading: consider the scenario initially flow~1 only passes through $SDM~1$ and $SDM~2$, later an operator decides to add an additional action to be performed on that flow and therefore $SDM~3$ is inserted into the service chain.
There can be many scenarios for such a need.
For example, a company can perform DPI, to block instant message packets (such as gtalk and Skype) only during the office hour.
Such temporary insertion/deletion of data plane instances should be seamless as no packet loss is allowed.
A simple solution could be to temporarily hold the packets of the flow, construct a new service chain, and switch to the new chain.
However, this approach adds significant delay to packets processing during the transition.
Instead, we prefer a bufferless solution.

\noindent\textbf{Control Plane~}
OpenFunction controller sends flow steering instructions to the Openflow controller.
We use OpenDaylight (ODL) \cite{ortiz2013software} as the OpenFlow controller that manages OpenFlow capable switches and routers.
OpenFunction controller runs as an OpenDaylight application.
It uses many base functions provided by ODL to learn about the network and control network elements.

In our testbed, we use OpenvSwitch \cite{ovs} as a switching element to steer traffic.
We use a quad core 3.0 GHz 	AMD Phenom(tm) II X4 945 Processor system to host the OpenFunction controller, the OpenFlow controller and all SDM CP processes.

\presub \subsection{Script Support} \postsub
Due to space limitation, we only explain how Netmap-Click and FPGA platforms works.
%
%
Netmap tool does not provide enough native packet libraries; therefore, we use Click as the packet processing pipeline.
%

%
%
The translated Click script of the IPSEC data plane specification (Algorithm~\ref{code:ipsec}) is shown in Algorithm~\ref{code:clickscript}.
One major challenge is the abstract parameters of actions in the specification: we either replace them with real parameters in Click format, or implement them as action attributes, and let the IPSEC CP process sets them in runtime.
Another challenge is that we need to add additional \emph{MarkMACHeader} and \emph{MarkIPHeader} elements, after the input device action, for Click specific requirements.

In FPGA box, both \sys~script and pseudo language are translated to the special C/C++ code: the specification is translated to the top level block (composed by a call graph among low level blocks), while each pseudo language based action definition is translated to a low level block.
The translation of data plane IPSEC specification (Algorithm~\ref{code:ipsec}) as shown in Algorithm~\ref{code:fpgatop}, requires
(1) defining the inter-link between functional actions (\ie, 5-20); and
(2) using direct function calls to represent actions (\eg, Line 21-29).
Note that in FPGA platform, we implement all parameters as action attributes, which are controlled by its corresponding CP process.

\vspace{-0.1in}
\presub \subsection{Language Support} \postsub
In Netmap platform, we translate each action pseudo program to an element class definition in Click.
Mostly, the generated codes are in the $simple\_action$ function.
Each special line in the pseudo language file is processed.
For example, Algorithm~\ref{code:clicksource}~is the translated code of generating a pointer to IP header length field (Algorithm~\ref{alg:SetIPChecksum}~Line 1).
Other codes are input from the pseudo program as is.
%


%
\sys~language translation in Xilinx FPGA is similar to the operations of that in Netmap.
The major difference is that: an additional C++ encapsulation class is required for each action class, to express the state transition semantics in FPGA hardware.
Due to space limitation, we omit the details of both FPGA and DPDK platforms.

\vspace{-0.1in}
\presub \subsection{In-Box Service Chain} 
\begin{figure}[tpb]
\centerline{\includegraphics[width=1\linewidth]{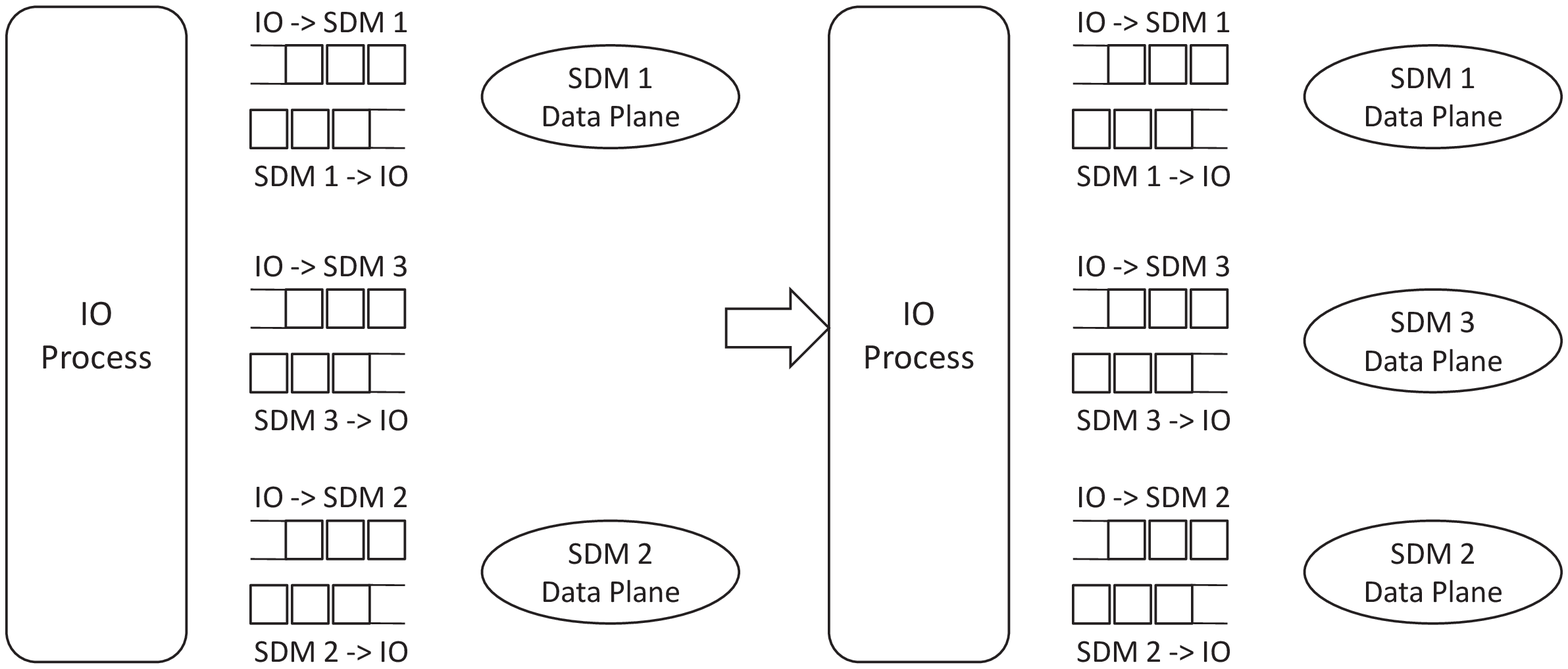}}
\caption{IO based service chain} \label{fig:ioproc}
\vspace{-0.2in}
\end{figure}

\postsub
In this part, we use DPDK to illustrate the provisioning of in-box service chain.
We use process group to realize the service chain and treat each SDM data plane instance as a separate process.
As shown in Figure \ref{fig:ioproc}, a single IO process acts as a central process: it sets up all memory structures for use and handles packet reception and transmission.
A pair of rings are setup between the IO process and a SDM process for packets exchange.
When a new SDM DP process starts, it discovers both the rings and packet memory locations via DPDK library support.
The IO process sends a packet's descriptor through the ring; a SDM data plane processes the packet and returns it via another ring.

The pair of rings is the key to realize dynamic load/unload.
Each packet's metadata carry the service chain information: an array records the indexes of ring pairs it needs to traverse in order, and an index pointer keeps the value of the next pair.
Considering the above-mentioned dynamic example, initially only $SDM~1$ and $SDM~2$ are traversed by a flow; later when the new $SDM~3$ data plane process starts, all subsequent packets' metadata change to the new service chain $SDM~1\rightarrow SDM~3\rightarrow SDM~2$; and as a result, packets traverse $SDM~3$ between $SDM~1$ and $SDM~2$.
Similarly, to unload a service chain, when $SDM~3$ is removed from the service chain, the metadata of all subsequent packets change to the new service chain $SDM~1\rightarrow SDM~2$ again; the controller waits for a while, before it cleans up the $SDM~3$ process from the system.
For performance consideration, we assume all processes of a group is running on a single CPU socket (possibly with multiple cores).
Note that multiple process group is supported, if we start different process groups in different sockets.

We also support this feature in the FPGA platform.
Unlike software, loading/uploading new data plane functions in programmable hardware needs direct support from hardware.
Partial Reconfiguration is a feature of modern FPGAs, which allows a subset of the logic fabric of a FPGA to be dynamically reconfigured while the remaining logic continues to
operate \cite{blodget2004partial}.
%

\vspace{-0.1in}
\presub \subsection{Developed SDMs} \postsub
We implement NAT and IPSec gateway as two stateless SDMs on all Netmap and FPGA platforms.
The NAT middlebox performs simple address translation; the mapping table is stored in its CP process and the mapping item is send to a DP process when the first packet of a flow comes in.
The IPSec gateway uses ESP tunnel model with AES-EBC encryption.
The control plain negotiates with the remote peer, and sends command to DP processes after security negotiation.

With Netmap, we also implement a simple stateful firewall and a stateful IPS.
The stateful firewall monitors all path-through TCP connections; it removes the allow rules in both directions whenever a RESET flag is detected.
The stateful IPS monitors FTP connections: some FTP commands are allowed only after the user authentication; otherwise an alert is raised.


%% file: alex/evaluation.tex
\presec \section{Evaluation} \label{sec:evaluation} \postsec
\subsection{Abstraction} \postsub
First, we test the  \sys~specification performance for the data plane abstraction of each SDM.
For each SDM, we use same set of \sys~specifications.
%
We test with both small size packets (60 bytes) and large size packets (1400 bytes).
Table~\ref{table:throughput} shows the throughput performance.
For non-stateful SDMs: NAT and IPSEC, DPDK box achieves higher throughput than Netmap-based box in almost all the scenarios.
This demonstrates that, even with the same OpenFunction abstraction implementation and the same hardware, the performance
actually relies on underlying systems.
In turn, the FPGA box is always better than DPDK.

\noindent\textbf{Key Takeaway 1~~} Our \sys~abstraction accommodates vendor differentiation with the same code across heterogeneous systems, which is beneficial for the NFV community.

\begin{table}[htb]
\caption{Throughput of \sys~middleboxes}\label{table:throughput}
\vspace{-0.15in}
\begin{center}
\scalebox{0.85}{
\begin{tabular}{|c|c|c|c|c|}
\hline & Netmap & DPDK & FPGA &  FPGA-C \\
\hline
\hline
NAT Small  & 0.14 Gbps & 1.84 & 1.64 Gbps &  0.24 Gbps\\
\hline
NAT Large &  0.20 Gbps & 9.96 & 9.34 Gbps  &  0.53 Gbps \\
\hline
IPSEC Small  & 0.02 Gbps & 1.2 & 2.61 Gbps &  0.04 Gbps\\
\hline
IPSEC Large &  0.17 Gbps & 2.8 & 9.35 Gbps  &  0.10 Gbps \\
\hline
\hline
FW Small  & 0.03 Gbps & N/A & N/A &  N/A\\
\hline
FW Large &  0.18 Gbps & N/A & N/A  &  N/A \\
\hline
IPS Small  & 0.02 Gbps & N/A & N/A &  N/A\\
\hline
IPS Large &  0.19 Gbps & N/A & N/A  &  N/A \\
\hline
\end{tabular}}
\end{center}
\end{table}

Next, we replace the \emph{SetIPChecksum} and \emph{SetTCPChecksum} actions in the specification with user defined versions (written in \sys~language and translated).
The performance overhead caused by user defined versions is negligible: for these software-based boxes, without special acceleration treatment, the C compiler performs most of the optimization work.
Also, for such an action, the efficiency of a translated version is similar to the implemented version.
%

For FPGA platform, the performance gap between optimized implementation and \sys~language generated version (\ie, FPGA-C) is significant (10x in this case)Table~\ref{table:throughput}.
The reason is that FPGA enables a  programmer to introduce several optimizations.
For example, for NAT function, the programmer may choose to calculate the checksum at line rate, without the need to receive complete packet.

\noindent\textbf{Key Takeaway 2~~} The performance of SDM data plane actions, if generated from \sys~pseudo language, is dominated by the nature of system.

Similarly, for two stateful SDMs: FW and IPS, the performance is comparable to non-stateful SDMs.

\noindent\textbf{Key Takeaway 3~~} \sys~abstraction has the potential to support various stateful firewalls.

\presub \subsection{Centralization} \postsub
In this section, we evaluate our algorithm by leveraging the traffic data collected from a production datacenter hosted by IBM global services.
To this end, we conduct performance evaluation from two perspectives.
First, we evaluate the impact of varying number of flows in the system on algorithm performance, and then the impact of threshold selected at line \ref{lin:threshold} in Algorithm \ref{alg:offline} on algorithm performance.

\begin{figure}
\centering
\includegraphics[width=3.4in]{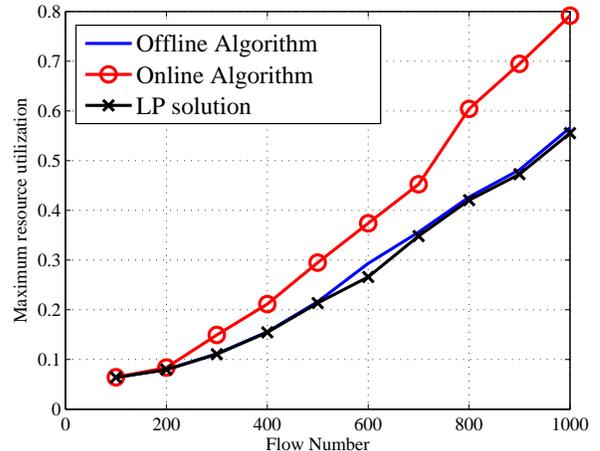}
\caption{Performance impacted by flow number.}\label{fig:schedule-flownum}
\vspace{-0.15in}
\end{figure}
\emph{Performance impact by varying the number of flows:}
In this section, we fix the box number to be 10, there are 6 SDMs and the required SDM number of each flow is evenly distributed in the range [1,5].
Each SDM has 2 or 3 versions implemented in the system.
In addition, we set the forbidden threshold to be 0.2, i.e. 20\% of the $Z_{k,i,n}^{l,j,m,f}$ and the fraction value according to the LP solution will be set to 0 in each iteration.
To see how the algorithm performance changes with the number of flows, we change the flow number from 100 to 1000, and the simulation results are shown in Fig. \ref{fig:schedule-flownum}.
We evaluate the performance of proposed online and offline algorithm and compare it to the LP solution, which treats all the flows as splittable, and hence the LP solution is a lower bound of our problem.
From Figure~\ref{fig:schedule-flownum}, we make two observations: \emph{i) the online algorithm achieves performance very close to LP solution, and ii) its performance margin from offline approach improves with increasing number of flows in the system.}

A single flow only needs very little system resource compared to the available system resources, and therefore, we can treat all the flows as splittable in the system's point of view.
As a result, the performance of offline algorithm is very close to the LP solution.

The performance of online algorithm deviates from the optimal solution as the number of flows increase in the system.
This is for the reason that the online algorithm is a myopic algorithm and optimizes the flows one by one, whenever a flow enters the system, it optimizes the maximum resource utilization based on the current system utilization but does not consider future flow arrivals.

\emph{The impact of threshold $\epsilon$:}
\begin{figure}
\centering
\includegraphics[width=3.4in]{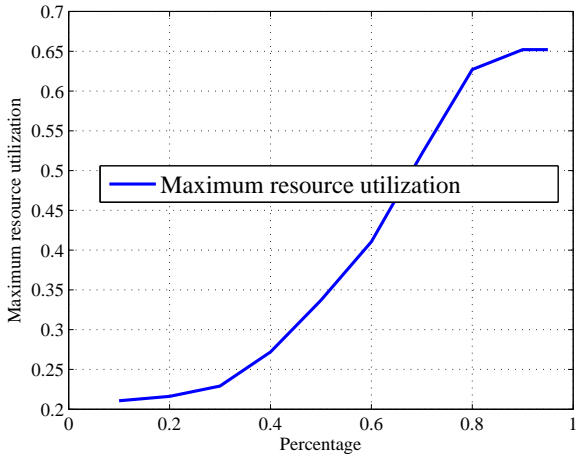}
\caption{Performance impacted by flow number.}\label{fig:schedule-percentage}
\vspace{-0.15in}
\end{figure}
In this section, we study how the forbidden parameter impacts the offline algorithm performance.
For this purpose, we inject 400 flows into the system and change the value of forbidden threshold from 0.1 to 0.95.

The simulation results, Fig. \ref{fig:schedule-percentage}, show that \emph{ for smaller thresholds, the offline algorithm performance degrades very slowly with increasing the value of forbidden threshold.
 However, when the threshold value exceeds a certain point, the algorithm performance degrades quickly.}
The reason is that in the LP solution, there are many $Z_{k,i,n}^{l,j,m,f}$ with very little fraction, and they can be set to 0 at one time without impacting the performance significantly.
%

In addition, as the value of the forbidden threshold continues to increase, the performance degradation slows down.
This happens because if we set too many  $Z_{k,i,n}^{l,j,m,f}$ as zero, the optimization space for the flows reduces significantly.
Therefore, forbidding more $Z_{k,i,n}^{l,j,m,f}$ cannot degrade the algorithm performance if the forbidden parameter is relatively large.

\noindent\textbf{Key Takeaway 4~~} \sys~abstraction has the potential to support various stateful firewalls.

Lastly, we consider two scenarios to evaluate OpenFunction data plane overhead:
(1) time it takes to start/stop a data plane process in the service chain; and
(2) additional path delays due to our chain enforcement inside a box.
In the first case, we are interested in understanding how much delay is incurred when a new service chain or SDM DP is added or removed from the network.
To evaluate, we use a small topology with 5 DPDK middleboxes and generate TCP flows between source and destination.
We have two DPDK boxes on path 1 and one DPDK box on path 2.
We initiate flows on path 1 and path 2 which initiates flow rule installation in the switches along the datapath and calculate datapath setup times.
Our evaluation shows that it takes less than 100\,msec to add a new datapath or service chain in the network.
%
%
In the second case, we are interested in understanding the additional delays caused by enforcing service chains in the MB's.
Our evaluations shows that it adds less than 20 usec per SDM DP in the service chain.


%
In this case, we are interested in knowing the responsiveness of the controller to adapt to dynamic network conditions and
we measure the time required to reconfigure the network in case of SDM DP failure or traffic overload in some of the service chains.
We consider same topology as before, with DPDK middleboxes, and use Iperf \cite{iperf} to generate TCP flows between source and destination.
We first initiate flows on $path 1$ and $path 2$ by install rules in the switches along the datapath and distribute load evenly across the two paths.
Next, we bring down the switch in $path 2$ and observe how quickly network converges to the stable state and diverts traffic to $path 1$.
Our evaluation shows that network takes less than 200\,msec to converge to a stable state and reconfigure flows to use other paths.
We also dynamically change the datapaths, to replicate the SDM insertion and deletion scenarios, in our topology to steer network traffic through different paths and our evaluation shows that the network quickly adapts to these changes without any drop in the flow throughput.

\noindent\textbf{Key Takeaway 5~~} \sys~abstraction is flexible and has light overhead.

%% file: alex/conclusion.tex
\presec \section{Conclusions} \label{sec:conclusions} \postsec
In this paper, we make three main contributions.
First, we propose the concept of Software-Defined Middleboxes to realize abstraction and centralization for middleboxes, which are complement existing SDN effort.
Second, we propose OpenFunction as an SDM reference architecture and a network function abstraction layer.
Third, we implemented a working OpenFunction system including one OpenFunction controller, three OpenFunction boxes, and four network functions including both stateful and stateless ones.
Our experimental results show that the middlebox functions implemented using our OpenFunction abstraction can achieve high performance and platform independence. 

%% file: alex/appendix.tex
\section*{Appendix}
\begin{algorithm} []
\caption{\textbf{IPSEC Script (one way)}}\label{code:ipsec}
@ FromDevice(1) fromdevice  \;
@ ExactMatch(PROTO\_IP,-) match \;
@ DecapHeader(14) decap\;
@ ESPEncap espencap\;
@ Aes(EBC) aes\;
@ IPsecEncap ipencap\;
@ ChangeSrcIP(IP\_SRC) srcip\;
@ ChangeDstIP(IP\_DST) dstip\;
@ SetIPChecksum  checksum\;
@ EncapHeader(MAC\_DEST, MAC\_SRC, PROTO\_IP) encap\;
@ ToDevice(2) todevice\;
\BlankLine
fromdevice 0 0 match;
match 0 0 decap \;
decap 0 0 espencap;
espencap 0 0 aes \;
aes 0 0 ipencap;
ipencap 0 0 srcip \;
srcip 0 0 dstip;
dstip 0 0 checksum \;
checksum 0 0 encap;
encap 0 0 todevice \;
match 1 0 discard;
\end{algorithm}

\begin{algorithm}
	\caption{\textbf{Translated Click IPSEC script}}\label{code:clickscript}
	in1 :: FromDevice(netmap:eth1, PROMISC true) \;
	out2 :: Queue(100)$\rightarrow$ToDevice(netmap:eth2) \;
	elementclass ExactMatch Classifier \;
	EM::ExactMatch(12/0800,-) \;
	\BlankLine
	in1$\rightarrow$EM$\rightarrow$DecapHeader(14)$\rightarrow$MyIPsecESPEncap \
	$\rightarrow$MyAes(1)$\rightarrow$IPsecEncap$\rightarrow$SetIPChecksum     \
	$\rightarrow$MarkIPHeader(OFFSET 0)$\rightarrow$ChangeSrcIP     \
	$\rightarrow$ChangeDstIP$\rightarrow$EncapHeader$\rightarrow$out2 \;
	\BlankLine
	EM[1]-$>$Discard \;
\end{algorithm}

\begin{algorithm}
	\caption{\textbf{Translated Click source}}\label{code:clicksource}
	nh\_data = p$\rightarrow$network\_header(); \
	\BlankLine
	unsigned char* pch = NULL \;
	pch = nh\_data \;
	pch = pch + 0 \;
	unsigned char* iphlen = pch \;
\end{algorithm}

\begin{algorithm}[t]
	\caption{Offline Resource Utilization Optimization}
	\label{alg:offline}
	\begin{algorithmic}[1]
		\REQUIRE Flow volume and service chain requirement, resource requirement for each SDM implementation
		\ENSURE The SDM path of each flow
		\STATE Formulate RUOP model according to the input
		\WHILE {not all $Z_{k,i,n}^{l,j,m,f} = 0 \text{or} 1$}
		\STATE Solve RUOP model
		\STATE Get a forbidden threshold $\epsilon$ \label{lin:threshold}
		\FOR{all $Z_{k,i,n}^{l,j,m,f}$}
		\IF{$Z_{k,i,n}^{l,j,m,f}<\epsilon$}
		\STATE Add one more constraint $Z_{k,i,n}^{l,j,m,f}=0$ into RUOP model
		\ENDIF
		\ENDFOR
		\ENDWHILE
		\RETURN $\{x_i^k\}$
	\end{algorithmic}
\end{algorithm}

\begin{algorithm}
\caption{\textbf{SetIPChecksum (User Defined)}}\label{alg:SetIPChecksum}
\KwIn{Packet \texttt{P}}
\KwOut{\texttt{P} with IP Checksum field set}
\BlankLine
@ FIELD \texttt{iphlen}   NETWORK 0  UINT8 \;
@ FIELD \texttt{ipcsum}   NETWORK 10 UINT8 \;
@ FIELD \texttt{ipheader} NETWORK 0  DATA \;
\BlankLine
unsigned char \texttt{hlen}\;
@ LOAD \texttt{iphlen} \texttt{hlen} \;
\texttt{hlen} = \texttt{hlen} \& 0x0f\;
\texttt{hlen} = \texttt{hlen} $\ll$ 2\;
\BlankLine
@ STORE \texttt{ipcsum} 0 \;
\BlankLine
unsigned int \texttt{sum}=0\;
unsigned short \texttt{hoffset}=0\;
unsigned short \texttt{tempshort}=0\;
\While {\texttt{hoffset} $<$ \texttt{hlen}}{
    \{
    \ @ LOAD \texttt{ipheader} \texttt{tempshort} \texttt{hoffset} UINT16 \;
    \texttt{hoffset} = \texttt{hoffset} + 2\;
    \texttt{sum} = \texttt{sum} + \texttt{tempshort}\;
    \}
}
\BlankLine
unsigned int \texttt{tempint} = 0\;
\texttt{tempint} = \texttt{sum} \& 0xffff0000\;
\While {\texttt{tempint} != 0}{
\{
\texttt{sum} = \texttt{sum} \& 0xffff\;
\texttt{tempint} = \texttt{tempint} $\gg$ 16\;
\texttt{sum} = \texttt{sum} + \texttt{tempint}\;
\texttt{tempint} = \texttt{sum} \& 0xffff0000\;
\}
}
\BlankLine
sum = $\sim$sum\;
@ STORE \texttt{ipcsum} \texttt{sum} \;
\end{algorithm}

\begin{algorithm}
\caption{\textbf{Translated Xilinx specification}}\label{code:fpgatop}
void ipsec(stream$<$axiWord$>$ \&inData, stream$<$axiWord$>$ \&outData) \{\\
\#pragma HLS dataflow interval=1 \\
\#pragma HLS INTERFACE port=inData axis  \\
\#pragma HLS INTERFACE port=outData axis \\
\BlankLine
	static stream$<$axiWord$>$ ippacket("ippacket");\\
	static stream$<$axiWord$>$ decap2espencap("decap2espencap");\\
    static stream$<$axiWord$>$ espencap2aes("espencap2aes");\\
    static stream$<$axiWord$>$ aes2ipencap("aes2ipencap");\\
    static stream$<$axiWord$>$ ipencap2changeSrcIP("ipencap2changeSrcIP");\\
	static stream$<$axiWord$>$ changeSrcIP2changeDstIP("changeSrcIP2changeDstIP");\\
	static stream$<$axiWord$>$ changeDstIP2ipchecksum("changeDstIP2ipchecksum");\\
    static stream$<$axiWord$>$ ipchecksum2macencap("ipchecksum2macencap");\\
\BlankLine
\#pragma HLS STREAM variable=ippacket 		depth=16    \\
\#pragma HLS STREAM variable=decap2espencap 		depth=16  \\
\#pragma HLS STREAM variable=espencap2aes 		depth=16  \\
\#pragma HLS STREAM variable=aes2ipencap 		depth=16  \\
\#pragma HLS STREAM variable=ipencap2changeSrcIP 	depth=16 \\
\#pragma HLS STREAM variable=changeSrcIP2changeDstIP 	depth=16 \\
\#pragma HLS STREAM variable=changeDstIP2ipchecksum depth=16 \\
\#pragma HLS STREAM variable=ipchecksum2macencap depth=16 \\
\BlankLine
	parser(inData, ippacket);\\
	decap(ippacket, decap2espencap);\\
    espencap(decap2espencap, espencap2aes);\\
    aes(espencap2aes,aes2ipencap);\\
    ipencap(aes2ipencap, ipencap2encap);\\
    encap(decap2encap, encap2changeSrcIP);\\
	changeSrcIP(encap2changeSrcIP, changeSrcIP2changeDstIP);\\
    changeDstIP(changeSrcIP2changeDstIP, changeDstIP2ipchecksum);\\
	ipchecksum(changeDstIP2ipchecksum, outData);\\
\}\\
\end{algorithm}